# Orientational migration of halide ion in lead halide perovskites

Ningbo Yi[1], Gang Li[1], Yuhan Wang[1], Shuai Wang[1], Kaiyang Wang[1], Qinghai Song[1,2,*], Shumin Xiao[1,2,#]

1. State Key Laboratory on Tunable laser Technology, Ministry of Industry and Information Technology Key Lab of Micro-Nano Optoelectronic Information System, Shenzhen Graduate School, Harbin Institute of Technology, Shenzhen, 518055, P. R. China.

2. Collaborative Innovation Center of Extreme Optics, Shanxi University, Taiyuan, 030006, Shanxi, P. R. China.

Email: # shumin.xiao@hit.edu.cn; * qinghai.song@hit.edu.cn

**Abstract:**

**With the rapid evolution of lead halide perovskite solar cells, there are considerable evidences that the mobile ions strongly impact their properties. However, except the activation energy, the other properties of ionic migration are mostly explored with first principle calculations. Here we investigate the migration of halide ion (vacancy) in lead bromide perovskite single crystalline microplates. In contrast to conventional studies with external bias voltages, we introduce "built-in" potentials into the microplates via low-energy electron beams and study the corresponding ionic migrations. By aligning the potentials along particular crystal planes, the reduction of $Pb^{2+}$ to $Pb^0$ has been observed at the un-irradiated center of each cell, demonstrating that the migration of halide ion has preferred directions for the first time. Our observations are consistent with the recent theoretical models and shall open up a new route to improve the performances of perovskite solar cells.**

Owing to their unique properties such as band-gap tunability[1], long charge diffusion length[2,3], low recombination rates[4], and photon recycling[5], solution-processed lead halide perovskite ($MAPbX_3$) have been promising materials for optoelectronic applications, especially for high-efficiency and cost-effective photovoltaic devices[6-9]. Within the past few years, the certified power conversion efficiency (PCE) of perovskite solar cells has been dramatically increased from a few percent to more than 22.7%[6,10]. Despite of these progresses, the perovskite solar cells usually show a different response between forward scan and backward scan during the current-voltage (J-V) characterization[11]. While the origin of the J-V hysteresis is still on debate, the movement of ionic species has been widely accepted to play an essential role in it[12,13]. Basically, the movements of ionic vacancies can generate a net charge accumulation in certain areas and produces a balancing internal counterfield. As the characteristic time of this process is of the same order of magnitude as the potential scanning time, this phenomenon can result in the well-known J-V hysteresis. Meanwhile, the ionic migration is also found to have important influences on the long-term stability of perovskite[14]. Consequently, the researches on the fundamental mechanism of ionic transport are essential for further improvements of perovskite optoelectronic devices[15,16]. In past few years, the movements of methyl ammonium ($MA^+$) and halide ions (vacancy) in lead halide perovskite have been theoretically simulated by several groups with different computational methodologies, producing a large range of activation energies ($E_A$)[15,16]. The theoretically computed $E_A$ ranges from 0.36 eV to 0.84 eV for $MA^+$ and from 0.08 eV to 0.6 eV for halide ions[12,16-23]. While the activation energy for halide ions is relatively smaller, it still overlaps largely with the $E_A$ for $MA^{+}$[24]. Therefore, in additional to the experimentally recorded $E_A$ (~ 0.17 eV - 0.6 eV)[16,21], more convincing evidences are highly desirable to determine the movable species in $MAPbX_3$ perovskites.

The recent first-principle calculations show that three vacancy transport mechanisms involve conventional hopping between neighbouring positions in $MAPbX_3$ perovskites[16]. As illustrated in Fig. 1, the $Pb^{2+}$ species migrate along the diagonal (<110> direction) of the cubic unit cell; the $X^-$ migrations follows an octahedron edge and the direction of displacement is the same as $Pb^{2+}$; the $MA^+$ species transport to a neighbouring vacant A-site cage. According to the recent simulation work, the $E_A$ for vacancy assisted diffusion of $Pb^{2+}$ (~2.31 eV) is usually much larger than the ones for $I^-$ and $MA^+$ and its influences on J-V hysteresis can thus be neglected. In this sense, the movable species ($I^-$ or $MA^+$) might be identified from the orientation of ionic migration in real experiment. However, the experimental determination of diffusion direction is quite challenging. While the migration follows particular direction in unit cells[14], in the absence of external field, the charged defects follow a random migration path. In case of solar cells or photodetectors, the migration is controlled by the external voltage or contacts and hard to determine the internal diffusion orientation. Here, in contrast to the previous studies, we introduce "built-in" potentials in single-crystalline $MAPbBr_3$ perovskite microplate and study the corresponding ionic migration. The $Br^-$ ions are found to have preferred orientations of migration.

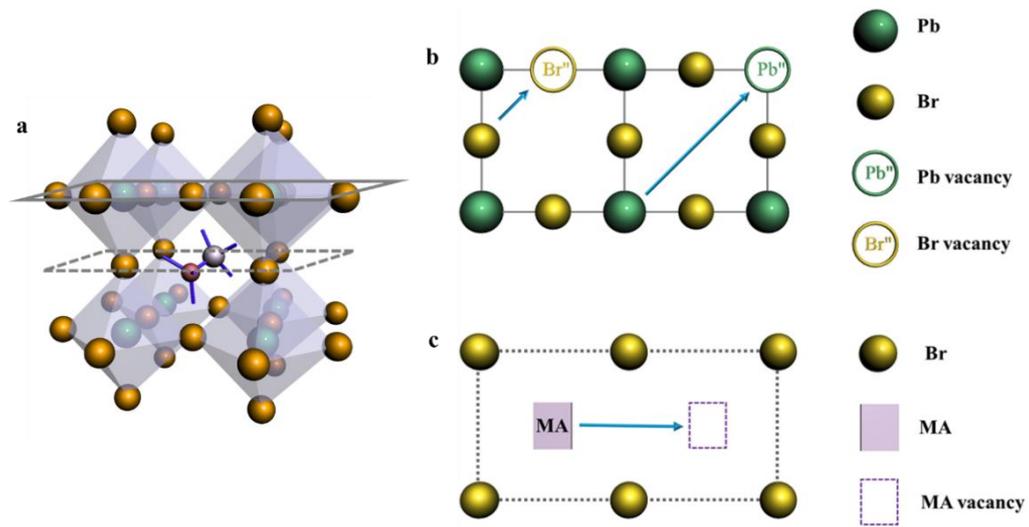

**Figure 1. The transport mechanisms in the MAPbBr$_3$ structure.** Schematic illustration of the three ionic transport mechanisms involving conventional vacancy hopping between neighbouring positions. (a) The perovskite structure of $MAPbBr_3$. (b) $Pb^{2+}$ transport along the diagonal direction <110>; $Br^-$ migration along an octahedron edge and its displacement follows the <110> too. (c) $MA^+$ migration into a neighbouring vacant A-site cage involving motion normal to the unit cell face composed of four bromide ions.

The perovskite crystals were synthesized with a one-step solution processed precipitation method (see methods, Figure S1 and S2). The synthesized microplates were irradiated with electron beam (E-beam) in an E-beam writer (Raith, E-line, 5-30 kV, Figure S3). Basically, there are two types of electron-specimen interactions, i.e. the knock-on damage and the radiolysis [25-27]. In case of low energy electron beam irradiation (LEEBR), the latter one is the dominant mechanism. Due to the weak bond between edge atoms and the surrounding lattices, the E-beam irradiation usually induces the desorption of a small fraction of bromide. As a result, some $Pb^{2+}$ are reduced to $Pb^0$ atoms, which aggregate at the surfaces and edges of perovskite microplate. Importantly, the LEEBR will generate negative charge accumulation[28] and induce a large number of bromide vacancies $V^*_{Br}$ inside the perovskite microplate[29,30]. Such kind of vacancies can function as potentials for the ions outside the irradiated region. The above radiolysis process has been verified with the X-ray photoelectron spectroscopy (XPS) and the atomic force microscope (AFM) image. After the LEEBR (20 kV, dose of 500 µC cm$^{-2}$, see details in methods), two additional peaks for $Pb^0$ appeared at 126 eV and 140.7 eV (see Figs.2a and 2d) and the intensity of peak for Br-Pb bands at 69.7 eV reduced obviously (see Figs. 2b and 2e). Meanwhile, numerous Pb nanoparticles have been observed on the top surface of the microplate (see Figs. 2c and 2f). All of these observations are consistent with previous reports and confirm the radiolysis process well[26].

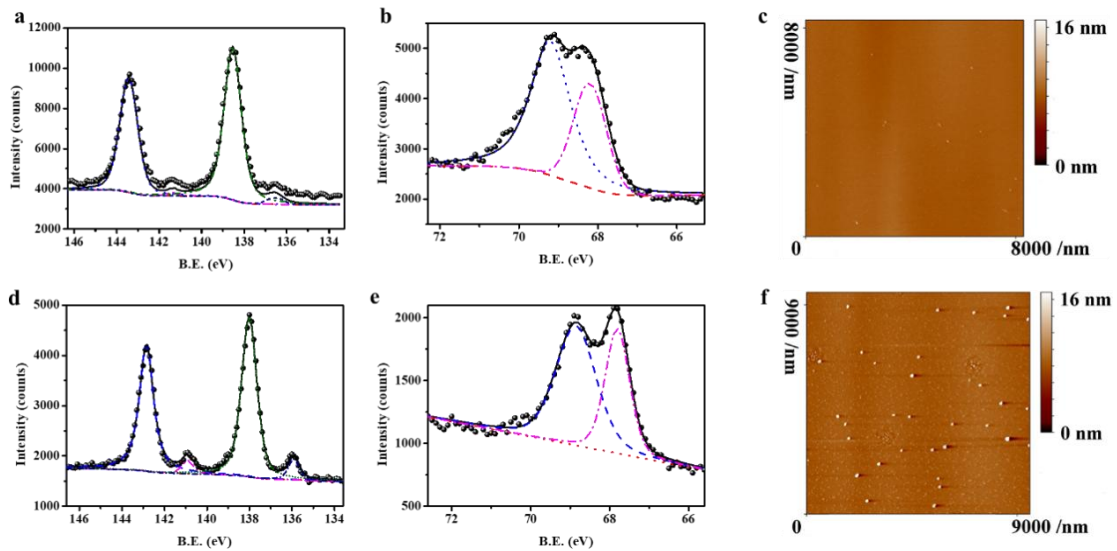

**Figure 2. The XPS measurements of MAPbBr$_3$ perovskite.** (a) Pb 4f and (b) Br 3d spectra of pristine MAPbBr$_3$ perovskite without E-beam irradiation. (c) The smooth surface of pristine crystal. (d) Pb 4f and (e) Br 3d spectra of MAPbBr$_3$ perovskite after E-beam irradiation. Here the acceleration voltage is 20 kV and the dose is 500 µC cm$^{-2}$. Two additional peaks of Pb 4f$_{5/2}$ and Pb 4f$_{7/2}$ appear in (d). (f) AFM image shows the formation of Pb nanoparticles at the top surface after the E-beam irradiation.

More interestingly, owing to the mature nanofabrication technique[26,31], the sizes and positions of irradiated areas have the possibility to be precisely controlled. As a result, periodic local potentials can be easily pattered in situ with an E-beam writer. The direction of the potentials is defined with an angle θ between the lattice direction and the crystal side (see the schematic picture in Fig. 3a). For a single potential, the Br$^-$ ions outside the irradiated region will be attracted and move towards it (see Fig.3b). During this process, the new vacancies generated by the ion migration will be quickly filled by the ions along the same movement direction and finally averaged out by the whole microplate. When four potentials are introduced (see points 1-4 in Fig. 3c), the situation becomes quite complicated. The vacancies at most positions such as point 5 can be filled by ions along at least two directions (marked by red arrows) and thus similar to the case with single potential. However, the situation at the special point, marked as 0 in Fig. 3c, is totally different. At this point, the Br$^-$ ions can be attracted by four potentials and move to points 1-4. Since the random ion migration without external potentials is much slower and is negligible, no additional Br$^-$ can replenish the vacancies and thus some Pb$^{2+}$ are reduced to Pb$^0$ atoms at point-0. As a result, even though the point has not been irradiated by the E-beam, additional nanoparticles are possibly formed on the top surface at the region around it.

    The newly generated nanoparticles can be detected by AFM. Thus they can be a very important clue to determine the directions of ion migrations in real experiments. Following the first-principle calculations, although the Br$^-$ ions migrate along an octahedron direction, the directions of displacements follow the <110> and its normal direction well[16]. For the MAPbBr$_3$ microplate, these directions are the diagonals of the square microplate (See Fig. 3a). Therefore, if the four potentials are aligned along the sides of microplates (<100> and <010> directions, see Fig. 3a), the potentials are introduced to attract Br$^-$ ions along the <110> and its orthogonal directions. As no additional Br$^-$ ions replenish the newly generated vacancies, the additional nanoparticles can be formed at the center of the square. In case that four potentials are aligned along the other direction, the ionic migration is quite difficult to follow particular directions. As a result, the Br ions at position-0 will not be depleted. Only four groups of nanoparticles can be formed at positions 1-4 and the center shall be very clean. Even if the theoretical predictions are not correct, the Br$^-$ ions can either migrate isotropically in all directions or transport along all of the crystal directions. In case of isotropic migration,

the additional nanoparticle group shall be observed at any $\theta$ value. For the latter one, the additional nanoparticles can be found at the other crystal directions such as <100> direction.

Based on the above discussion, the MAPbBr$_3$ microplate was irradiated by E-beam with an accelerated voltage of 20 kV, dose of 500 μC cm$^{-2}$. The diameters of the irradiated circles and the side length of square are designed as 1.5 μm and 4 μm, respectively. And the direction of square was designed along the side of microplate with $\theta = 0°$. The morphology of the microplate has been examined by AFM after the irradiation. The result is plotted in Fig. 3d. Four groups of nanoparticles can be clearly seen at the irradiation regions, consistent with the previous reports well[26]. Interestingly, an additional group of nanoparticles appeared at the center of the square, where the MAPbBr$_3$ perovskite was not irradiated by the E-beam at all. In additional to the morphology, the current-AFM (c-AFM) has also been used to investigate the irradiated MAPbBr$_3$ microplate (see methods). As shown in Fig.3f, the irradiated position such as position-1 has much stronger photocurrent than the current at un-irradiated position-5 (background value). Following the previous report, the increased photocurrent is attributed to the formation of Pb nanoparticles[26]. This information can be further confirmed by the photocurrent at position-0. Even though it was also not irradiated, the photocurrent was similar to the one at position-1 and much larger than the background value at position-5. Therefore, the additional group of nanoparticles at position-0 can confirm the ion migration along the <110> and its orthogonal directions in the MAPbBr$_3$ microplate. This is consistent with the first-principle calculations well and confirms that the Br$^-$ instead of the MA$^+$ are the movable species.

In order to exclude the other possibilities of ion migrations, we have increased the angle $\theta$ of the "built-in" potentials and studied the morphology of top surface after the LEEBR. Figure 3e shows the morphology image of MAPbBr$_3$ perovskite microplate with the angle of potential $\theta = 45°$. In this case, the diagonals of square correspond to the sides of the perovskite microplate (<100> and <010> directions). Four groups of Pb nanoparticles can still be seen at the irradiation positions. However, the center of the square is very clean and no additional Pb nanoparticles are observed. Similar AFM images have also been observed at the other $\theta$ values except $\theta = 0°$ (see SI). These experimental observations show that the ionic transport is quite difficult along the crystal directions such as <100> and <010>, simply excluding the other two possibilities mentioned above. Therefore, the theoretical model in previous reports has been experimentally verified for the first time that the ionic migration has preferred directions.

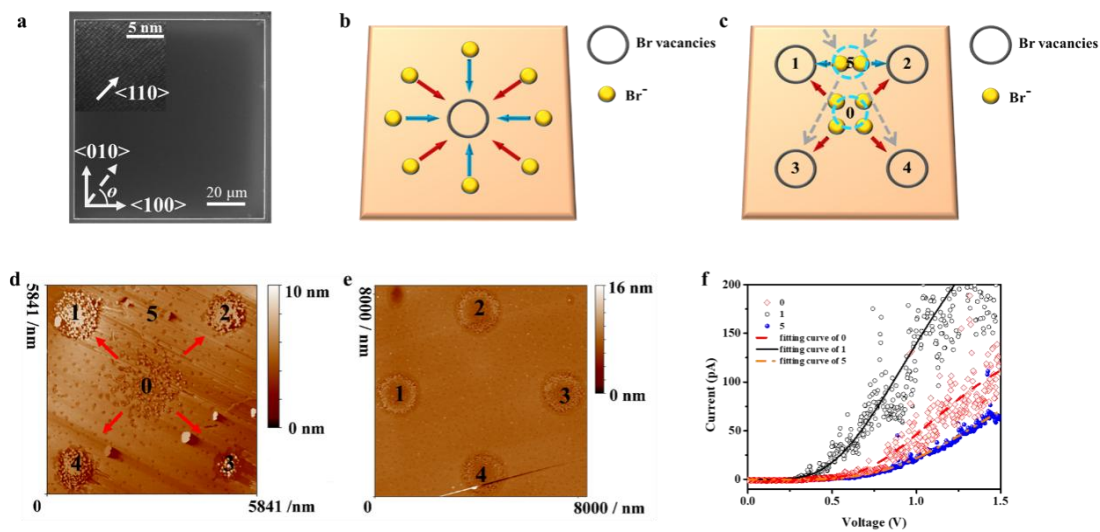

**Figure 3. The observation of orientational ion migration in MAPbBr$_3$ perovskite.** (a) The schematic picture of the introduction of local potentials into a MAPbBr$_3$ perovskite microplate. The potentials are aligned with an angle θ. (b) The illustration of ion migration with a single potential. (c) Four local potentials are built in the microplate. The ion migration at position-5 and position-0 are marked by arrows. Different from position-5, the Br$^-$ at position-0 will be depleted and thus generate the well-known reduction of Pb$^{2+}$ to Pb$^0$. (d) The AFM image of the MAPbBr$_3$ perovskite microplate after the introduction of four "built-in" potentials. At the center of the square, an additional group of Pb nanoparticles can be observed. Here the direction of square is aligned at θ = 0. (e) A similar AFM image of the MAPbBr$_3$ perovskite microplate after the introduction of four "built-in" potentials. As the θ is changed to 45°, the reduction of Pb$^{2+}$ to Pb$^0$ can only be seen at the irradiated positions. (f) The photocurrents of E-beam irradiated microplate in (d) measured by the current-AFM. The current at the un-irradiated position-0 is close to the irradiated position and much higher than the background current at position-5.

We have also examined the dependence of the additional Pb nanoparticle group on the other parameters such as acceleration voltage. By keeping the direction θ, the morphologies of irradiated microplate are studied as a function of acceleration voltage, dose, and the side length of square. All the experimental results have been summarized in SI. The additional Pb nanoparticle group at the un-irradiated central region has been always observed when the acceleration voltage is 8 kV (dose 200 μC cm$^{-2}$), 10 kV (dose 250 μC cm$^{-2}$), 15kV (dose 375 μC cm$^{-2}$), or 20 kV (dose 500 μC cm$^{-2}$) in Figure S4. Similar phenomena also holds true when the separation distance between two nearby potentials changes from 4 μm to 6 μm, 7 μm and 8 μm in Figure S5. However, once the direction θ changes from 0$^o$, the additional Pb nanoparticle groups disappear in in Figure S6. As a result, we can conclude that the orientational ionic transport is quite generic and robust in lead halide perovskite.

In summary, we have experimentally studied the orientational ion migrations in single crystalline MAPbBr$_3$ microplates. By introducing well defined local potentials into the microplate, we experimentally observed that the ion migration along <110> and its orthogonal directions are much easier than the others. These experimental observations are consistent with the first-principle calculations very well and verified the preferred ionic migration directions for the very first time. This research will be important for understanding of ionic transport in lead halide perovskite and further optimizing the lead halide perovskite based optoelectronic devices. Meanwhile, the "built-in" potentials can also be applied in many other materials as an additional way to characterize or change their local characteristics.

**Methods**:

**The synthesis of MAPbBr$_3$ microplates:** we have experimentally synthesized CH$_3$NH$_3$PbBr$_3$ perovskite microplates and studied their changes in optical properties under electron beam irradiation. The CH$_3$NH$_3$PbBr$_3$ perovskite microplates were synthesized with the solution processed one-step precipitation method. Basically, the CH$_3$NH$_3$PbBr$_3$ precursor was prepared by dissolving PbBr$_2$ and CH$_3$NH$_3$Br with a 1:1 molar ratio in N, N-dimethylsulfoxide to give a concentration of 0.05 m, followed by a 4 h stirring at 60 ℃. The indium tin oxide (ITO) glasses were cleaned by sonication in acetone, isopropanol, and deionized water for 15 min, respectively. Then the CH$_3$NH$_3$PbBr$_3$ precursor was drop casted on to the ITO glass, which was located in atmosphere of dichloromethane in a beaker. The beaker was sealed with tetrafluoroethylene films. After a reaction time of ~24 h, the perovskite microplates with relatively large dimensions have been formed on the ITO glass.

**The characterization:** The morphological images were obtained with a scanning probe microscope (CSPM5500, Benyuan Nano instruments Co. Ltd) in tapping mode. The photocurrent distribution on the surface of crystal was obtained with an atom force microscope (Bruker MultiMode 8) in contact mode with a bias of 1.5 V. XPS was performed with ESCALAB 250Xi (Thermo Fisher), equipped with a monochromatized Al Kα X-ray source (1486.6 eV). Energy calibration was performed fixing the C-C component of C 1s spectrum at 284.8 eV. Relative atomic percentages of different species were computed from high-resolution spectra, fitted with XPSPEAK software. XRD was performed in Rigaku D/Max2500, equipped with a X-ray source of Cu kα1.

# Supporting Information：

# Orientational migration of halide ion in lead halide perovskites

Ningbo Yi[1], Gang Li[1], Yuhan Wang[1], Shuai Wang[1], Kaiyang Wang[1], Qinghai Song[1,2,\*], Shumin Xiao[1,2,#]

3. State Key Laboratory on Tunable laser Technology, Ministry of Industry and Information Technology Key Lab of Micro-Nano Optoelectronic Information System, Shenzhen Graduate School, Harbin Institute of Technology, Shenzhen, 518055, P. R. China.

4. Collaborative Innovation Center of Extreme Optics, Shanxi University, Taiyuan, 030006, Shanxi, P. R. China.

Email: # shumin.xiao@hit.edu.cn; * qinghai.song@hit.edu.cn

To demonstrate that ions migration is induced by ions vacancies in crystal, not the electron beam with high energy, the patterns on the surface of perovskite are written with different accelerating voltages, 8 kV, 10 kV, 15 kV and 20 kV, and corresponding doses of 250, 375 and 500 μC cm$^{-2}$ in Figure S4. From the morphology analysis, the ions migration shows an independence on accelerating voltages. This is because that the Br escapes from the surface in the form of $Br_2$ upon interaction between electron beams with perovskite according to Knotek–Feibelman (K-F) mechanism. And then, the vacancies of $Br^-$ ions forms on the local sites to break the balance of crystal structure, which induces the hopping migration of Br ions in neighbor to ions vacancies along a path with the lowest barrier. Therefore, this kind of ions migration observed in AFM images is not in relation to the electron beams, but just to the number and arrangement of ions vacancies created by electron beams.

Similar phenomena also holds true when the separation distance between two potentials changes from 4 μm to 6 μm, 7 μm and 8 μm in Figure S5. Herein, the patterns was irradiated by electron beam lithography with an accelerated voltage of 20 kV, dose of 500 μC cm$^{-2}$. The designed circle radius is 0.75 μm.

Several directions ($\theta$) were investigated for the influences on ions migration. As shown in Figure S6, once the direction $\theta$ changes from 0$^o$, the additional Pb nanoparticle groups disappear at the center of four patterns. This is an importance demonstration for the direction of ions migration in perovskite crystal.

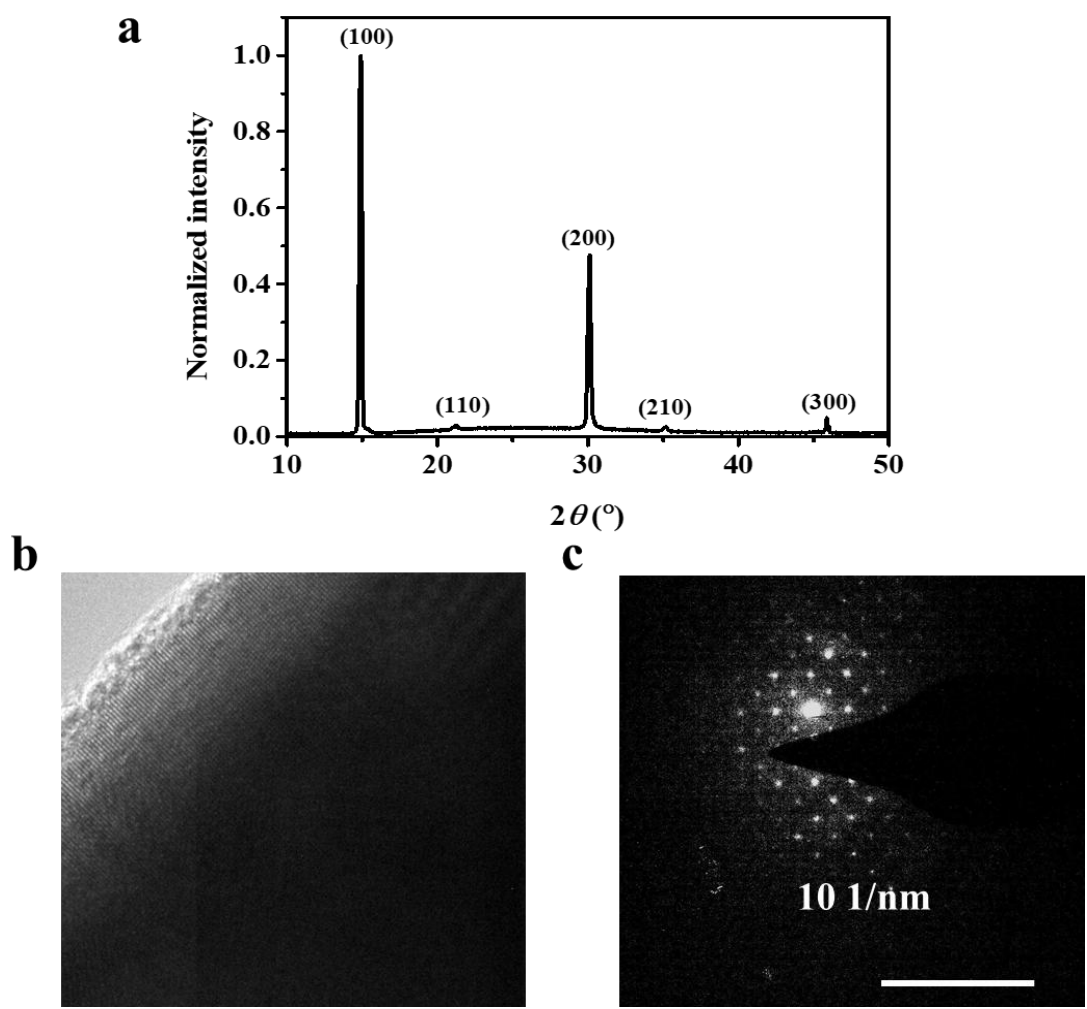

Figure S1 (a) The XRD spectrum, (b) TEM image and (c) SAED image indicate a single crystal of perovskite, which is used in this study.

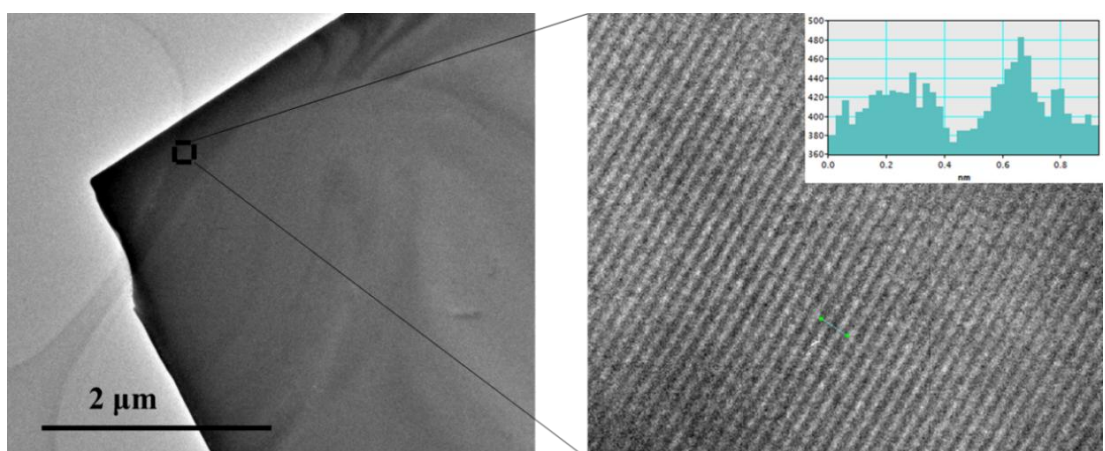

Figure S2 The high-resolution TEM images of single crystal. The distance of two plate indicated by line is ~0.437 nm, corresponding to the spacing of (110).

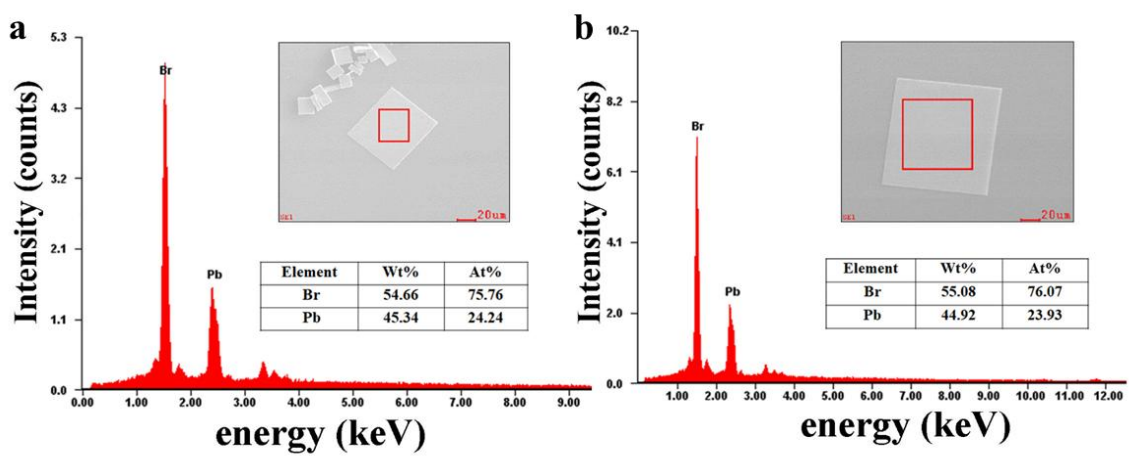

Figure S3 The EDS analysis of $CH_3NH_3PbBr_3$. (a) The pristine sample. (b) The sample after electron beam irradiation with a dose of 500 μC cm$^{-2}$ under a voltage of 20 kV.

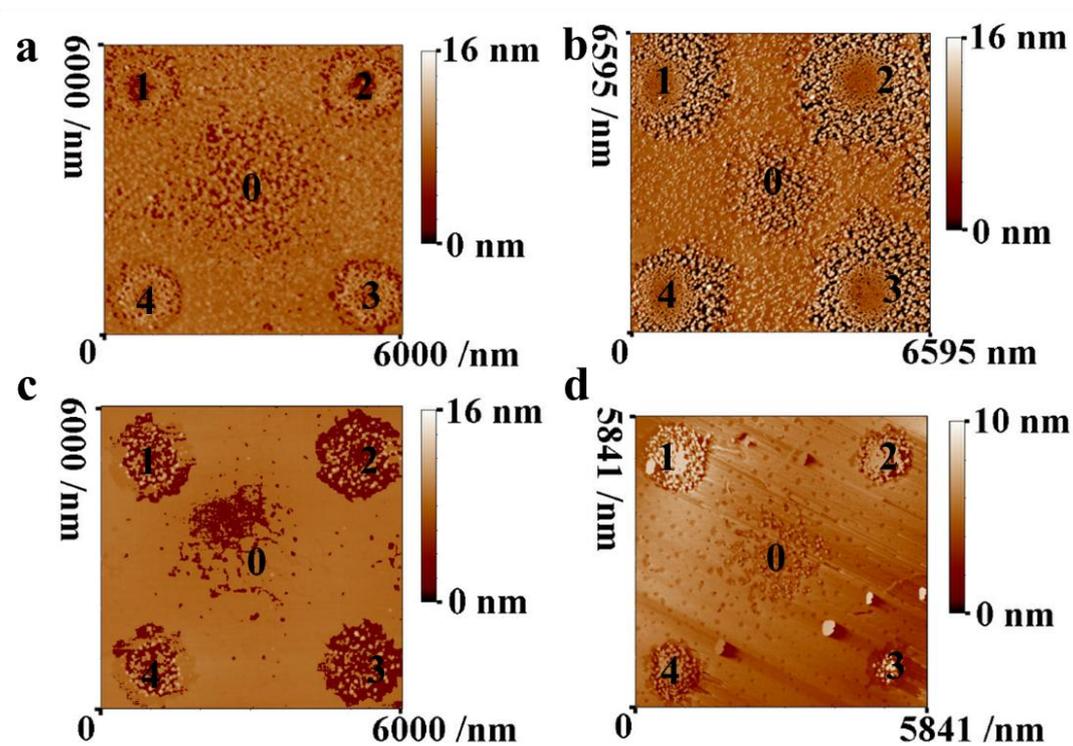

Figure S4 The additional Pb nanoparticle group at the un-irradiated central region has been always observed when the acceleration voltage is 8 kV(a, dose of 200 μC cm$^{-2}$), 10 kV (b, dose of 250 μC cm$^{-2}$), 15kV (c, dose of 375 μC cm$^{-2}$), and 20 kV (d, dose of 500 μC cm$^{-2}$) by keeping the direction $\theta$ of 0°.

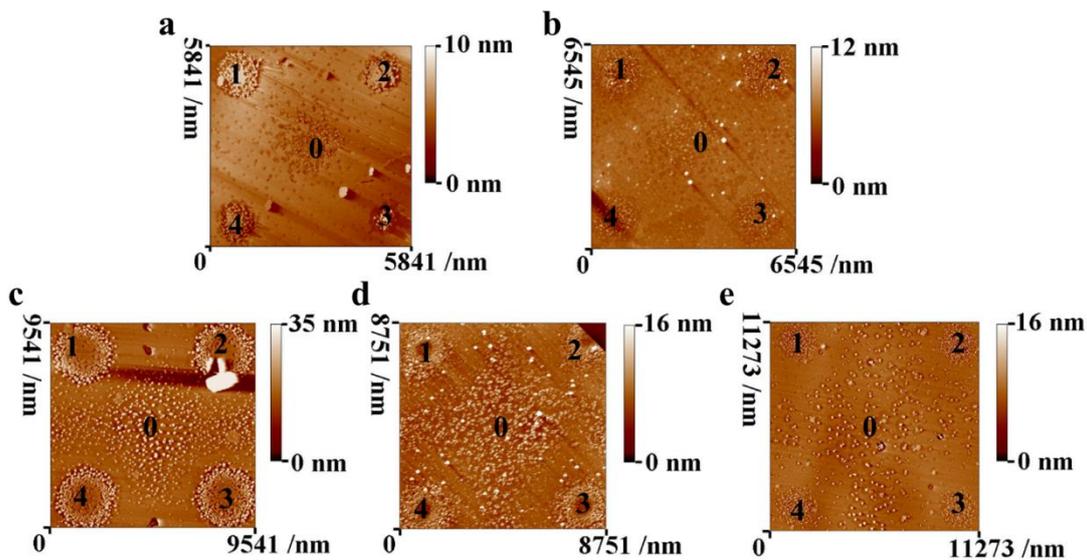

Figure S5 The additional Pb nanoparticle group at the un-irradiated central region has been always observed with the separation distance between two potentials changes: (a) 4 μm, (b) 5 μm, (c) 6 μm, (d) 7 μm and (e) 8 μm.

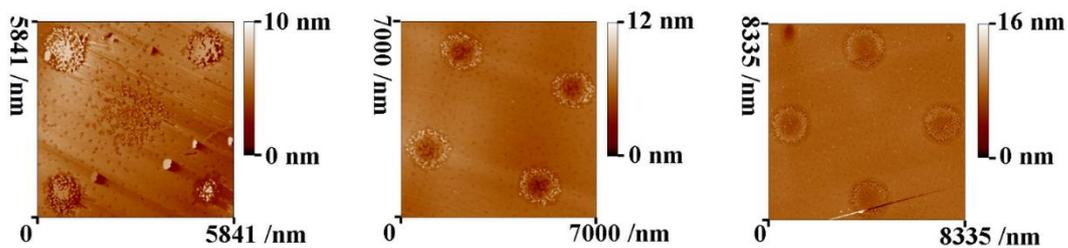

Figure S6 The dependence for ions migration on the direction (*θ*) of potential formation on the surface of perovskite crystal. (a) *θ* = 0°.(b) *θ* ≠ 0°. (c) *θ* = 45°.